\documentstyle[12pt]{article}
\newcommand{\beqn}{\begin{eqnarray}}
\newcommand{\eeqn}{\end{eqnarray}}
\newcommand{\be}{\begin{equation}}
\newcommand{\ee}{\end{equation}}
\newcommand{\ba}{\begin{array}}
\newcommand{\ea}{\end{array}}
\newcommand{\R}{{\rm\bf R}}
\newcommand{\C}{{\rm\bf C}}
\newcommand{\pa}{\partial}
\newcommand{\re}{\ref}
\newcommand{\ci}{\cite}
\newcommand{\la}{\label}
\newcommand{\bfr}{\begin{flushright}}
\newcommand{\efr}{\end{flushright}}
\newcommand{\bfl}{\begin{flushleft}}
\newcommand{\efl}{\end{flushleft}}
\newcommand{\fr}{\frac}
\newcommand{\ov}{\overline}

\newcommand{\si}{\sigma}
\newcommand{\al}{\alpha}

\newcommand{\ds}{\displaystyle}
\newcommand{\ve}{\varepsilon}

\newcommand{\de}{\delta}
\newcommand{\om}{\omega}
\newcommand{\Om}{\Omega}

\newcommand{\br}{|\kern-.25em|\kern-.25em|}

\newcommand{\De}{\Delta}

\begin{document}
\renewcommand{\theequation}{\thesection.\arabic{equation}}
\newtheorem{theorem}{Theorem}[section]
\renewcommand{\thetheorem}{\arabic{section}.\arabic{theorem}}
\newtheorem{definition}[theorem]{Definition}
\newtheorem{deflem}[theorem]{Definition and Lemma}
\newtheorem{lemma}[theorem]{Lemma}
\newtheorem{example}[theorem]{Example}
\newtheorem{remark}[theorem]{Remark}
\newtheorem{remarks}[theorem]{Remarks}
\newtheorem{cor}[theorem]{Corollary}
\newtheorem{pro}[theorem]{Proposition}
\newcommand{\HH}{{\rm\bf H}}
\newcommand{\pr}{\prime}
\def\N{{\rm I\kern-.1567em N}}                              
\def\No{\N_0}                                               
\def\R{{\rm I\kern-.1567em R}}                              
\def\C{{\rm C\kern-4.7pt                                    
\vrule height 7.7pt width 0.4pt depth -0.5pt \phantom {.}}\,}
\def\Z{{\sf Z\kern-4.5pt Z}}                                
\def\P{{\rm P\kern-6.0pt P}}                                
\def\n#1{\vert #1 \vert}                                    
\def\nn#1{\Vert #1 \Vert}                                   
\def\Re {{\rm Re\, }}                                       
\def\Im {{\rm Im\,}}                                        
\newcommand{\loc}{\scriptsize loc}
\newcommand{\const}{\mathop{\rm const}\nolimits}
\newcommand{\curl}{\mathop{\rm curl}\nolimits}
\newcommand{\nul}{\mathop{\rm nul}\nolimits}
\newcommand{\tr}{\mathop{\rm tr}\nolimits}
\newcommand{\diver}{\mathop{\rm div}\nolimits}
\newcommand{\supp}{\mathop{\rm supp}\nolimits}
\newcommand{\ow}{\overrightarrow}

\begin{titlepage}
\hspace{5cm} 
 {\em Russian J. Math. Physics} {\bf 10} (2003), no.4, 399-410.
\vspace{5mm} 
\begin{center}
{\Large\bf
 On the Convergence to a Statistical Equilibrium \bigskip\\
for the Dirac  Equation}
\end{center}
\vspace{2cm}
 \begin{center}
{\large T.V. Dudnikova}
\footnote{Supported partly by the
research grants of DFG (436 RUS 113/615/0-1) and
of RFBR (01-01-04002) and by
the Austrian Science Foundation
(FWF) Project (P16105-N05) }\\
{\small\it
M.V.Keldysh Institute\\
of Applied Mathematics RAS\\
 Moscow 125047, Russia\\
e-mail:~dudnik@elsite.ru}
\bigskip\\
{\large A.I. Komech$^{1,}\!\!$
\footnote{Supported partly by
the Austrian Science Foundation
(FWF) START Project (Y-137-TEC)}
}\\
{\small\it Wolfgang-Pauli Institute\\
c/o Institute of Mathematics, Vienna University \\
1090 Vienna, Austria\\
 e-mail:~komech@mat.univie.ac.at}
\bigskip\\
{\large N.J. Mauser}\\
{\small\it Wolfgang-Pauli Institute\\
c/o Institute of Mathematics, Vienna University \\
1090 Vienna, Austria\\
 e-mail:~mauser@courant.nyu.edu}
\end{center}
 \vspace{1cm}

 \begin{abstract}
We consider  the Dirac equation in $\R^3$
with constant coefficients and  study the distribution $\mu_t$
 of the random solution at  time $t\in\R$.
It is assumed that the initial measure $\mu_0$ has zero mean,
a translation-invariant covariance, and 
 finite mean charge density. We also assume
that $\mu_0$ satisfies a  mixing condition of
Rosenblatt- or Ibragimov-Linnik-type.
 The main result is the convergence of $\mu_t$ to
 a Gaussian measure as $t\to\infty$.
The proof uses the study of long time asymptotics
 of the  solution and
S.N. Bernstein's ``room-corridor''  method.

 \end{abstract}
\end{titlepage}

 \section{Introduction}
This paper can be regarded as a continuation of our
papers \ci{DKKS}-\ci{DKM1}
 concerning the analysis of long-time
 convergence to an equilibrium distribution
for hyperbolic  partial differential equations
 and harmonic crystals.
Here we develop the  analysis for
 the  Dirac equation 
\beqn
\left\{\ba{l}
\dot\psi(x,t)=
\left[-\al\cdot\nabla-i\beta m\right]
\psi(x,t),\,\,\,\,x\in\R^3,\\
\psi(x,0)=\psi_0(x),
\ea
\right.\la{21'}
\eeqn
where   $\nabla=(\pa_1,\pa_2,\pa_3)$,
$\pa_k=\ds\pa/\pa x_k$, $m>0$,
$\al=(\al_1,\al_2,\al_3)$,
$\al_k$ and $\beta$ are $4\times 4$ Dirac matrices
(see (\ref{ba}), (\ref{sigma})).
The solution
$\psi(x,t)\in \C^4$ for $(x,t)\in\R^4$.

It is assumed that the initial data $\psi_0(x)$ are given
by a random element of the 
 function space ${\cal H}\equiv H^0_{loc}(\R^3)$
of   states with finite local energy,
see Definition \ref{d1.1} below.
The distribution of $\psi_0$ is a zero-mean
probability measure $\mu_0$  satisfying some additional
assumptions, see Conditions {\bf S1-S3} below.
Denote by  $\mu_t$, $t\in\R$, the measure
on  ${\cal H}$ giving the  distribution of
the random solution  $\psi(t)$ of  problem (\re{21'}).
We identify the complex and real spaces
$\C^4\equiv\R^8$,
and  $\otimes$ stands for the tensor product of real vectors.
The correlation functions of the initial measure
are supposed to be translation-invariant,
\be\la{1.9'}
Q_0(x,y):= E\Big(\psi_0(x)\otimes\psi_0(y)\Big)=
q_0(x-y),\,\,\,x,y\in\R^3.
\ee
We also assume that the
initial mean charge density is finite,
\be\la{med}
e_0:=E \vert \psi_0(x)\vert^2=
\tr q_0(0)<\infty,\,\,\,\,x\in\R^3.
\ee
Finally, assume that the  measure $\mu_0$ satisfies
 a mixing
condition of a Rosenblatt- or Ibragimov-Linnik
type, which means that
\be\la{mix}
\psi_0(x)\,\,\,\,   and \, \, \,\,\psi_0(y)
\,\,\,\,  are\,\,\,\, asymptotically\,\,\,\, independent\,\, \,\,
 as \,\, \,\,
|x-y|\to\infty.
\ee
Our main result gives the (weak) convergence
of $\mu_t$ to a limit measure $\mu_\infty$,
\be\la{1.8i}
\mu_t \rightharpoondown
\mu_\infty,\,\,\,\, t\to \infty,
\ee
which is an equilibrium Gaussian measure on
${\cal H}$. A similar convergence holds for $t\to-\infty$
because our system is time-reversible.
Explicit formulas (\re{1.13})
for the correlation functions of  $\mu_\infty$
are given.
\medskip

To prove  the convergence (\re{1.8i}) we follow the strategy
of \ci{DKKS}-\ci{DKM1}. There are  three steps.\\
{\bf I.}
The family of measures
 $\mu_t$, $t\geq 0$, is weakly
compact in an appropriate Fr\'echet space.\\
{\bf II.}
The correlation functions converge to a limit,
\be\la{corf}
Q_t(x,y)\equiv
\int \psi(x)
\otimes\psi(y)\,\mu_t(\psi)
\to Q_\infty(x,y),\,\,\,\,t\to\infty.
\ee
{\bf III.}
The characteristic functionals converge to a Gaussian functional,
\be\la{2.6i}
 \hat \mu_t(\phi):
 =  \int e^{i\langle\psi,\phi\rangle}\mu_t(d\psi)
\rightarrow \exp\{-\fr{1}{2}{\cal Q}_\infty (\phi,\phi)\},
\,\,\,\,t\to\infty.
 \ee
Here
$\langle\cdot,\cdot\rangle$ stands for
 a real  scalar product in $L^2$,
${\cal Q}_\infty$ for a  quadratic form
with the  integral kernel
$Q_\infty(x,y)$, and
$\phi$  for an arbitrary element of  the dual space.
\medskip

Property {\bf I} follows from the Prokhorov Compactness
Theorem by a method used in \ci{VF}.
Namely, we  first establish a uniform bound for
the mean local charge with respect to the
  measure $\mu_t$, $t\geq 0$.
Then the Prokhorov condition follows from the Sobolev
embedding theorem by Chebyshev's inequality.
Property  {\bf II} is derived
from an analysis of oscillatory
integrals arising in the Fourier transform.
However, the Fourier transform by itself is insufficient
to prove Property {\bf III}.
We derive it by using  an explicit representation
of the solution in the coordinate space
with the help of the Bernstein' ``room-corridor''
technique
by a method of  \ci{DKKS}-\ci{DKM1}.
The method gives a representation
 of the solution
as a sum of weakly dependent random variables.
Then (\ref{1.8i}) follows from the
Ibragimov-Linnik central limit theorem
under a Lindeberg-type condition.
We sketch the proofs by
using the technique of \ci{DKKS}.

The paper is organized as follows.
The  main result is stated in Section 2.
The compactness (Property {\bf I})
 is established in Section 3,
the  convergence (\re{corf}) in Section 4,
and  the convergence (\re{2.6i}) in Sections 5.

\setcounter{equation}{0}
 \section{Main results}
Let us describe our results more precisely.
\subsection{Notation}

We assume that the initial date $\psi_0$  in (\ref{21'})
is complex-valued vector function belonging
 to  the phase space ${\cal H}$.
 \begin{definition}                 \la{d1.1}
Denote by ${\cal H} \equiv H_{\rm loc}^0(\R^3,\C^4)$
  the Fr\'echet space
of complex-valued functions $\psi(x)$,
 endowed with local energy seminorms
\be\la{4}
\Vert \psi\Vert^2_{0,R}\equiv
\int\limits_{|x|<R}
|\psi(x)|^2\,dx
  <\infty,   ~~ \forall R>0.
\ee
 \end{definition}

 \begin{pro}    \la{p1.1}
(i) For any $\psi_0 \in {\cal H}$
 there exists  a unique solution
$\psi(\cdot,t)\in C(\R,\,{\cal H})$
 to Cauchy problem (\re{21'}).\\
 (ii) For any  $t\in \R$, the
operator $U(t):\psi_0\mapsto  \psi(\cdot,t)$
 is continuous in ${\cal H} $.
 \end{pro}

Proposition \ref{p1.1} follows from
\ci[Thms. V.3.1, V.3.2]{Mikh})
because the speed of propagation for Eq. (\ref{21'}) is finite.

Let us choose a function
$\zeta(x)\in C_0^\infty(\R^3)$such that $\zeta(0)\ne 0$.
Denote by $H^s_{\rm loc}(\R^3),$ $s\in \R,$  the local Sobolev spaces,
i.e., the Fr\'echet spaces
of distributions $u\in D'(\R^3)$ with the finite seminorms
\be\la{not}
\Vert u\Vert _{s,R}:= \Vert\Lambda^s\Big(\zeta(x/R)u\Big)\Vert_{L^2(\R^3)},
\ee
where $\Lambda^s v:=F^{-1}_{k\to x}(\langle k\rangle^s\hat v(k))$,
$\langle k\rangle:=\sqrt{|k|^2+1}$, and $\hat v:=F v$ is the Fourier transform
of a tempered distribution $v$. For $\phi\in C_0^\infty(\R^3)$
 write $F\phi ( k)= \ds\int e^{i k\cdot x} \phi(x) dx.$
Note that  the space  $H_{\rm loc}^{s}(\R^3)$
for $s=0$ agrees with Definition \ref{d1.1}.
 \begin{definition}\la{d1.2}
For $s\in\R$,  write
 $ {\cal H}^{s}\equiv H_{\rm loc}^{s}(\R^3). $
 \end{definition}

Using the standard technique of pseudodifferential operators
 and Sobolev's embedding theorem (see, e.g., \ci{H3}),
 one can prove
 that  ${\cal H}={\cal H}^0\subset {\cal H}^{-\ve }$ for every $\ve>0$,
and the embedding  is compact.

\subsection{Random solution. Convergence to equilibrium}
Let $(\Om,\Sigma,P)$ be a probability space
with expectation $E$ and let 
${\cal B}({\cal H})$ be the Borel $\sigma$-algebra
of ${\cal H}$.
Assume that $\psi_0=\psi_0(\om,\cdot)$ in (\re{21'})
is a measurable random function
with values in $({\cal H},\,{\cal B}({\cal H}))$.
In other words, $(\om,x)\mapsto \psi_0(\om,x)$
is a measurable  mapping
$\Om\times\R^3\to\C^4$ with respect to the
(completed) $\sigma$-algebras
$\Sigma\times{\cal B}(\R^3)$ and ${\cal B}(\C^4)$.
Then,  by virtue of Proposition \re{p1.1},
$\psi(t)=U(t) \psi_0$ is again a measurable  random
function with values in
$({\cal H},{\cal B}({\cal H}))$.
Denote by $\mu_0(d\psi_0)$ the Borel probability measure
on ${\cal H}$ giving
the distribution of  $\psi_0$.
Without loss of generality,
 we can assume that $(\Om,\Sigma,P)=
({\cal H},{\cal B}({\cal H}),\mu_0)$
and $\psi_0(\om,x)=\om(x)$ for
$\mu_0(d\om)\times dx$-almost all points
$(\om,x)\in{\cal H}\times\R^3$.
\begin{definition}
Let $\mu_t$ be the probability measure on ${\cal H}$
 giving the distribution of $Y(t)$,
\begin{eqnarray}\la{1.6}
\mu_t(B) = \mu_0(U(-t)B),\,\,\,\,
\forall B\in {\cal B}({\cal H}),
\,\,\,   t\in \R.
\eeqn
\end{definition}

Our main objective is to derive
 the weak convergence of the measures $\mu_t$
in the Fr\'echet space
 ${\cal H}^{-\ve }$ for each  $\ve>0$,
 \be\la{1.8}
 \mu_t\,\buildrel {\hspace{2mm}{\cal H}^{-\ve }}\over
 {- \hspace{-2mm} \rightharpoondown }
 \, \mu_\infty {\rm ~~ as ~~}t\to \infty,
 \ee
 where $\mu_\infty$ is some Borel probability measure
on the space ${\cal H}$.
 This means the convergence
 \be\la{1.8'}
 \int f(\psi)\mu_t(d\psi)\rightarrow
 \int f(\psi)\mu_\infty(d\psi), ~~~~t\to \infty,
 \ee
 for any bounded  continuous functional $f(\psi)$
 on  ${\cal H}^{-\ve }$.

Set
 ${\cal R}\psi\equiv (\Re \psi,\Im \psi)=
\{\Re\psi_1,\dots,\Re\psi_4,\Im\psi_1,\dots,\Im\psi_4\}$
for $\psi= (\psi_1,\dots\psi_4)\in \C^4$,
and   denote by  ${\cal R}^j\psi$ 
 $j$th component of the vector ${\cal R}\psi$, $j=1,...,8$.
The brackets $(\cdot ,\cdot)$
 mean  the  inner product in the real  Hilbert spaces
 $L^2\equiv L^2(\R^3)$,  in $L^2\otimes \R^N$, or in
some their   extensions.
For $\psi(x),\phi(x)\in L^2(\R^3,\C^4)$, write
\be\la{1.5'}
\langle\psi,\phi\rangle:=
 ({\cal R}\psi,{\cal R}\phi)=
\sum\limits_{j=1}^8({\cal R}^j\psi,{\cal R}^j\phi).
\ee
\begin{definition}
The correlation functions of the measure $\mu_t$ are
defined by
\be\la{qd}
  Q_t^{ij}(x,y)\equiv E\Big({\cal R}^i\psi(x){\cal R}^j\psi(y)\Big)
\quad\mbox{for almost all }\,\, x,y\in\R^3,~~i,j=1,...,8,
\ee
provided that the expectations in the RHS are finite.
\end{definition}

Denote  by $D$ the space of complex-valued functions
in $C_0^\infty(\R^3)$ and write ${\cal D}:=[D]^4$.
For a Borel probability  measure $\mu$ on  ${\cal H}$,
 denote by $\hat\mu$
the characteristic functional (the Fourier transform)
$$
\hat \mu(\phi )  \equiv
 \int\exp(i\langle\psi,\phi \rangle)\,\mu(d\psi),\,\,\,
 \phi\in  {\cal D}\quad (\mbox{see }~(\ref{1.5'})).
$$
A  measure $\mu$ is said to be {\it Gaussian}
 (with zero expectation) if
its characteristic functional is of  the form
$$
\ds\hat{\mu}(\phi ) =  \ds \exp\Big\{-\fr{1}{2}
 {\cal Q}(\phi , \phi )\Big\},\,\,\,\phi \in {\cal D},
$$
where ${\cal Q}$ is a real nonnegative
quadratic form on ${\cal D}$.
A measure $\mu$ is said to be {\it translation-invariant} if
$$
\mu(T_h B)= \mu(B),\,\,\,\,\, B\in{\cal B}({\cal H}),
\,\,\,\, h\in\R^3,
$$
where $T_h \psi(x)= \psi(x-h)$, $x\in\R^3$.

\subsection{Mixing condition}
Let $O(r)$ be the set of all pairs of open bounded subsets
${\cal A}, {\cal B}\subset \R^3$ at the distance
dist$({\cal A},\,{\cal B})\geq r$, and let
$\sigma ({\cal A})$ be the $\sigma$-algebra in
 ${\cal H}$ generated by the
linear functionals
$\psi\mapsto\, \langle\psi,\phi\rangle$,
where  $\phi\in  {\cal D}$
with $ \supp \phi \subset {\cal A}$.
Define the
Ibragimov-Linnik mixing coefficient
of a probability measure $\mu_0$ on  ${\cal H}$
 by  formula (cf. \ci[Def. 17.2.2]{IL})
\be\la{7}
\varphi(r)\equiv
\sup_{({\cal A},{\cal B})\in O(r)} \sup_{
\ba{c} A\in\si({\cal A}),B\in\si({\cal B})\\ \mu_0(B)>0\ea}
\fr{| \mu_0(A\cap B) - \mu_0(A)\mu_0(B)|}{ \mu_0(B)}.
\ee
\begin{definition}
We say that the measure $\mu_0$ satisfies the  strong, uniform
Ibragimov-Linnik mixing condition if
\be\la{1.11}
\varphi(r)\to 0\quad{\rm as}\quad r\to\infty.
\ee
\end{definition}
We specify the rate of  decay of $\varphi$ below 
(see Condition {\bf S3}).


\subsection{Main assumptions and results}

We assume that the measure $\mu_0$
has the following properties {\bf S0--S3}:
\bigskip\\
{\bf S0.}
$\mu_0$ has zero expectation value,
$E\psi_0(x)  \equiv  0,\,\,\,x\in\R^3$.\\
{\bf S1.}
 $\mu_0$ has translation-invariant correlation functions,
\be\la{q0}
  Q_0^{ij}(x,y)\equiv E\Big({\cal R}^i\psi(x){\cal R}^j\psi(y)\Big)=q^{ij}_0(x-y)
\,\,\,\,\mbox{for almost all }\,\, x,y\in\R^3,~~i,j=1,...,8.
\ee
{\bf S2.}   $\mu_0$ has a finite
 mean charge density, i.e., Eq. (\re{med}) holds.\\
{\bf S3.}
 $\mu_0$ satisfies the strong uniform
Ibragimov-Linnik mixing condition, with
 \be\la{1.12}
\int _0^\infty r^{2}\varphi^{1/2}(r)dr <\infty.
 \ee
The standard form of the Dirac matrices
$\al_k$ and  $\beta$
(in $2\times 2$ blocks) is
\be\la{ba}
\beta= \left(
\ba{ll}
I & 0\\
0 & -I\\
\ea  \right),\quad
\al_k=
\left(
\ba{ll}
0 & \sigma_k\\
\sigma_k & 0\\
\ea  \right)\quad (k=1,2,3),
\ee
where  $I$ denotes the  unit matrix and
\be\la{sigma}
\sigma_1=  \left(
\ba{ll}
0 & 1\\
1 & 0\\
\ea         \right),\quad
\sigma_2=   \left(
\ba{ll}
0 & -i\\
i & 0\\
\ea         \right),\quad
\sigma_3=   \left(
\ba{ll}
1 & 0\\
0 & -1\\
\ea         \right).
\ee
Introduce the following $8\times 8$
real valued matrices (in $4\times 4$ blocks)
\be\la{matr}
\Lambda_1= \left(
\ba{ll}
\alpha_1 & 0\\
0 & \alpha_1\\
\ea  \right),~~
\Lambda_2= \left(
\ba{ll}
0&i\alpha_2\\
- i\alpha_2 &0\\
\ea  \right),~~
\Lambda_3= \left(
\ba{ll}
\alpha_3 & 0\\
0 & \alpha_3\\
\ea  \right),~~
\Lambda_0= \left(
\ba{ll}
0&-\beta\\
 \beta&0\\
\ea  \right).
\ee
Note that
by (\ref{ba}) and (\ref{sigma})
we have
$$
i\al_2= \left(
\ba{ll}
0 & i\sigma_2 \\
i\sigma_2 & 0\\
\ea  \right),~~ \mbox{where  }\, i\sigma_2=
\left(
\ba{ll}
0 & 1\\
-1 & 0\\
\ea  \right).
$$
Moreover,
$\Lambda_k^T=\Lambda_k$,
$k=1,2,3$, $\Lambda_0^T=-\Lambda_0$.
Write
\be\la{LP}
{\Lambda}=(\Lambda_1,\Lambda_2,\Lambda_3),\quad
 P(\nabla)={\Lambda}\cdot\nabla+m\Lambda_0.
\ee
For almost all $x,y\in\R^3$, introduce
 the  matrix-valued function
\be\la{Q}
Q_{\infty}(x,y)\equiv
\Big(Q_{\infty}^{ij}(x,y)\Big)_{i,j=1,\dots,8}
=\Big(q_\infty^{ij}(x-y)\Big)_{i,j=1,\dots,8}.
\ee
Here
\beqn \la{1.13}
 \hat q_\infty (k)
=\frac{1}{2}\hat q_0(k)+\frac{1}{2}
\hat{\cal P}(k)P(-ik)\hat q_0(k) P^T(ik),
\eeqn
where $\hat{\cal P}(k)=1/(k^2+m^2)$,
and $\hat q_0(k)$  is
the Fourier transform of the correlation matrix
of the measure $\mu_0$ (see (\ref{q0})).
Since $P^T(ik)=-P(-ik)$, we have, formally, 
\beqn \la{1.13'}
 q_\infty (z) =\frac{1}{2}q_0(z)-\frac{1}{2}
{\cal P}* P(\nabla) q_0(z)
P(\stackrel{\leftarrow}\nabla),
\eeqn
where ${\cal P} (z)=e^{-m|z|}/(4\pi|z|)$
is the fundamental solution for the
operator $-\De+m^2$, and
   $*$ stands fors the convolution of distributions.
We  show below that
 $\hat q_0\in L^2\equiv L^2(\R^3)$ (cf ({4.7})).
 Hence, $\hat q_\infty(k)\in L^2$ 
by (\ref{1.13}), and the convolution in (\ref{1.13'})
also belongs to  $L^2$.
\medskip

Denote by ${\cal Q}_{\infty}$ a real quadratic form
on $L^2$ defined by
$$
{\cal Q}_\infty (\phi,\phi)\equiv
(  Q_\infty(x,y),
 {\cal R}\phi(x)\otimes {\cal R}\phi(y))=
\sum\limits_{i,j=1}^8
\int\limits_{\R^3\times\R^3}
\!  Q_\infty^{ij}(x,y)
{\cal R}^i\phi(x) {\cal R}^j\phi(y)\,dxdy.
$$
The form ${\cal Q}_{\infty}$ is continuous in $L^2$ 
because $\hat q_\infty(k)$ is bounded by Corollary \ref{coro}.
\medskip\\
{\bf Theorem A.}
{\it    Let  $m>0$, and  let {\bf S0--S3} hold.
 Then \\
(i) the convergence in (\re{1.8}) holds for any $\ve>0$.\\
(ii) The limit measure
$ \mu_\infty $ is a Gaussian equilibrium
measure on ${\cal H}$.\\
(iii) The  characteristic functional of $\mu_\infty$
is of the form
$$
\ds\hat { \mu}_\infty (\phi ) = \exp
\Big\{-\fr{1}{2}
{\cal  Q}_\infty ( \phi,  \phi)\Big\},\,\,\,
\phi \in {\cal D}.
$$
}
 Theorem A  can be derived
 from Propositions  \re{l2.1} and \re{l2.2}
 given below 
 by using the same arguments as in
\ci[Theorem XII.5.2]{VF}.
\begin{pro}\la{l2.1}
 The family of  measures $\{\mu_t, t\in\R\}$
 is weakly compact in the space
 ${\cal H}^{-\ve}$ for any $\ve>0$.
\end{pro}

\begin{pro}\la{l2.2}
For any $\phi\in {\cal D}$,
 \beqn\la{2.6}
\hat \mu_t(\phi ) \equiv
\int\exp\{i\langle\psi,\phi\rangle\}\,\mu_t(d\psi)
=E\exp\{i\langle U(t)\psi,\phi\rangle\}
 \rightarrow \exp\Big\{-\fr{1}{2}{\cal  Q}_\infty
(\phi,\phi)\Big\},\,\,t\to\infty.\,
 \eeqn
\end{pro}
Propositions \re{l2.1} and  \re{l2.2}  are proved in
Sections 3 and 4-5, respectively.

\subsection{Remark on various mixing conditions
for the initial measure}

  We use the {strong uniform}
Ibragimov-Linnik mixing condition for the simplicity of our
 presentation.
The {\it uniform} Rosenblatt mixing condition
\ci{Ros}  with a higher
degree $>2$  in the bound (\re{med})
is also sufficient. In this case we assume that
 there exists a $\delta$, $\de >0$,
such that
$
\sup\limits_{x\in\R^3}
E\vert \psi_0(x)\vert^{2+\de}<\infty.
$
Then condition (\re{1.12}) requires the following
 modification:
$$\ds \int _0^\infty r\al^{p}(r)dr <\infty,
\quad p=\min(\de/(2+\de), 1/2),
$$
where $\al(r)$ is the  Rosenblatt
mixing coefficient  defined
as in  (\re{7}), but without the denominator $\mu_0(B)$.
The statements of Theorem A and their
proofs remain essentially unchanged.
\setcounter{equation}{0}
   \section{Compactness of measures}
\subsection{Fundamental solution of the Dirac operator}
One can easily check that $\al_k$ and $\beta$
are Hermitian symmetric matrices
satisfying the anti-commutation relations
$$
\left\{
\ba{ll}
\al^*_k=\al_k,&\beta^*=\beta,\\
\al_k\al_l+\al_l\al_k=2\delta_{kl}I,&
\al_k\beta+\beta\al_k=0,
\ea
\right.
$$
( $\de_{kl}$ is Kronecker's delta).
Therefore,
$$
\left(\partial_t+\al\cdot\nabla+i\beta m\right)
\left(\partial_t-\al\cdot\nabla-i\beta m\right)=
(\partial^2_t -\triangle+m^2)I.
$$
Then we can construct
a fundamental solution ${\cal E}(x,t)$ of the Dirac operator,
 i.e., a solution of the  equation
$$
\left(\partial_t+\al\cdot\nabla+i\beta m\right) 
{\cal E}(x,t)=\delta(x,t)I,\quad {\cal E}(x,t)=0\,\,\,\,\,
\mbox{ for }\,\,\, t<0,
$$
 of the form
\be\la{23}
{\cal E}(x,t)=
\left(\partial_t-\al\cdot\nabla-i\beta m\right) 
E(x,t),
\ee
where  $ E(x,t)\equiv  E_t(x)$ 
is a fundamental solution for
the Klein-Gordon operator
$(\partial^2_t -\triangle+m^2)$,
and $E$ vanishes for $t<0$.

\begin{remark}\la{r2.1}
The function $ E_t(x)$ is given by
\be\la{fs}
 E_t(x)=
F^{-1}_{k\to x}\frac{\sin \om t}{\om},
\quad \omega\equiv\omega(k)\equiv \ds\sqrt{|k|^2+m^2}.
\ee
Then, by the Paley-Wiener
Theorem (see, e.g., \ci[Theorem II.2.5.1]{EKS}),
the function of $ E_t(\cdot)$ is supported by the ball
$|x|\le t$.
\end{remark}

Denote by $U(t),$ $t\in\R,$ the dynamical group for 
problem (\ref{21'}).
Then  $U(t)$ is a convolution operator given by
\be\la{25}
\psi(x,t)=U(t) \psi_0={\cal E}(\cdot,t)*\psi_0=
\left(\partial_t-\al\cdot\nabla-i\beta m\right)
E_t(\cdot)*\psi_0.
\ee
The convolution  exists because the
distribution ${\cal E}(\cdot,t)$ is compactly supported
by (\ref{23}) and by Remark \ref{r2.1}.

\subsection{Local  estimates}
\begin{pro} \la{p2.1}
For every $\psi_0\in{\cal H}$   and $R>0$,
\be\la{26}
\Vert U(t)\psi_0\Vert_{0,R}\le C
\Vert\psi_0\Vert_{0,R+t},~~t\in\R,
\ee
where $C<\infty$ does not depend on $R$ and $t$.
\end{pro}
{\bf Proof.}
In the Fourier transform,
the solution $\psi(x,t)$ of the
 Cauchy problem (\ref{21'}) reads as
$$
\hat \psi(k,t)=\hat{\cal E}(k,t)\hat\psi_0(k)=
\Big[\cos\om t-(\al\cdot (-ik)+i\beta m)
\frac{\sin\om t}{\om}\Big]\hat\psi_0(k)
$$
 by (\ref{23}) and (\ref{fs}).
Then, for $\psi_0\in L^2$, 
\beqn\la{estl2}
\Vert\psi(\cdot,t)\Vert_{L^2}=
\Vert\hat\psi(\cdot,t)\Vert_{L^2}
\le C\Vert\hat\psi_0(\cdot)\Vert_{L^2}=
C\Vert\psi_0(\cdot)\Vert_{L^2}.
\eeqn
Let us  consider  $\psi_0\in {\cal H}$.
Introduce the function $\psi_{0}^*(x)$ equal to
$\psi_0(x)$ for $|x|\le R+t$ and to $0$ otherwise.
Denote by $\psi(x,t)$ (by $\psi^*(x,t)$) the solution 
of the Cauchy problem (\ref{21'})
with the initial data  $\psi_{0}(x)$ 
( $\psi_{0}^*(x)$, respectively).
Note that  $\psi(x,t)=\psi^*(x,t)$ for $|x|\le R$.
Therefore, relation (\ref{estl2}) implies
\beqn
\Vert\psi(\cdot,t)\Vert_{R}=\Vert\psi^*(\cdot,t)\Vert_{R}
\le C \Vert\psi_{0}^*(\cdot)\Vert_{L^2}=
C \Vert\psi_{0}(\cdot)\Vert_{R+t}.\nonumber
\,\,\,\,\,\,\,\,\,\Box
\eeqn


\subsection{Proof of compactness}
Proposition \re{l2.1} follows
from the estimate (\ref{p3.1}) below by
using the Prokhorov Theorem
 \cite[Lemma II.3.1]{VF}, as in the proof of
 \ci[Thm. XII.5.2]{VF}.
\begin{pro}\la{p3.1}
Let the conditions of Theorem~A hold.
Then, for any positive $R$, there exists a constant $C(R)>0$
such that
\be\la{41}
\sup_{t\ge 0}E \Vert U(t)\psi_0\Vert^2_{0,R}\le C(R)<\infty.
\ee
\end{pro}
{\bf Proof.}
Let us write
 \be \la{3.2}
e_t(x):= E|\psi(x,t)|^2,\,\,\,\, x\in\R^3.
\ee
The mathematical expectation is finite  for almost
every $x$ by (\ref{26})
and by the Fubini theorem. Moreover, 
$e_t(x)=e_t$  for almost every  $x\in\R^3$ by 
Condition {\bf S1}. Hence, it follows from
the Fubini theorem,  (\re{26}) and Condition
{\bf S2} that
\be\la{43}
E \Vert U(t)\psi_0\Vert^2_{0,R}\equiv
e_t|B_R|\le C E \Vert\psi_0\Vert^2_{0,R+|t|}\equiv
C e_0|B_{R+|t|}|,~~t\in\R.
\ee
 Here $B_R$ is the ball $| x| \le R$ in $\R^3$, and
 $| B_R| $ is the volume of this ball.
As $R\rightarrow \infty$, we see from (\re{43}) that
 $e_t\le Ce_0$.
Thus, $$
E\Vert U(t) \psi_0\Vert^2_{0,R}=
e_t|B_R|\le Ce_0|B_R|<\infty.\,\,\,\,\,\,\,\,\,\,\,\Box
$$

\setcounter{equation}{0}
\section{Convergence of correlation functions}
We prove the convergence
of the correlation functions for the  measures $\mu_t$.
This implies   Proposition \ref{l2.2}  in the case of
 Gaussian measures $\mu_0$.
It follows from condition {\bf S1} that
\be\la{4.0}
Q_t^{ij}(x,y)=q^{ij}_t(x-y),\quad x,y\in\R^3,
\ee
  for $i,j=1,\dots,8$.
\begin{pro}\la{p4.1}
The correlation functions $q_t^{ij}(z)$, $i,j=1,\dots,8$,
converge for any $z\in\R^3$,
\be\la{4.4}
q_t^{ij}(z)\to q_{\infty}^{ij}(z),\quad t\to\infty,
\ee
where the functions $q_{\infty}^{ij}(z)$ are defined
in (\ref{1.13}).
\end{pro}
{\bf Proof.}
Using the notation (\ref{matr}) and (\ref{LP}),
by (\ref{25})  we obtain
$$
{\cal R}\psi(x,t)=\Big(\pa_t -
P(\nabla)\Big)E_t*{\cal R}\psi_0.
$$
Then,  by (\ref{fs}) and (\ref{LP}), 
 the Fourier transform of 
 the solution  to Cauchy problem (\ref{21'})
 becomes
\be\la{hatpsi}
\widehat{{\cal R}\psi}(k,t)=
\hat {\cal G}_t(k)\widehat{{\cal R}\psi_0}(k),\,\,\,\,\,\,
\mbox{where }\,\,
\hat {\cal G}_t(k):=\cos \omega t-P(-ik)
\frac{\sin\omega t}{\omega}.
\ee
The translation invariance (\ref{q0}) implies that
\be
E(\widehat{{\cal R}\psi_0}(k)\otimes\widehat{{\cal R}\psi_0}(k'))=
F_{x\to k,y\to k'}q_0(x-y)=(2\pi)^3\de(k+k')\hat q_0(k).
\ee
Further, (\ref{hatpsi}) gives
\be
E(\widehat{{\cal R}\psi}(k,t)\otimes\widehat{{\cal R}\psi}(k',t))=(2\pi)^3\de(k+k')\hat {\cal G}_t(k)\hat q_0(k)\hat {\cal G}^*_t(k).
\ee
Therefore,
by the inverse Fourier transform  we obtain
\beqn
q_t(x-y)&=&E({\cal R}\psi(x,t)\otimes {\cal R}\psi(y,t))=
F^{-1}_{k\to(x-y)}\hat {\cal G}_t(k)\hat q_0(k)\hat {\cal G}^*_t(k)\nonumber\\
&=&(2\pi)^{-3}\int e^{-ik(x-y)}\Big(\cos \omega t-P(-ik)
\frac{\sin\omega t}{\omega}\Big)\hat q_0(k)
\Big(\cos \omega t-P^T(ik)
\frac{\sin\omega t}{\omega}\Big)\,dk\nonumber\\
&=&(2\pi)^{-3}
\int e^{-ik(x-y)}\Bigl[ \frac{1+\cos 2\omega t}{2}
\hat q_0(k)-\frac{\sin 2\omega t}{2\omega}
\Bigl( \hat q_0(k)P^T(ik)+
  P(-ik)\hat q_0(k)\Bigr)
\nonumber\\
&&+\frac{1-\cos 2\omega t}{2\omega^2}
P(-ik)\hat q_0(k)P^T(ik) \Bigr]\,dk.
\la{4.6}
\eeqn
To prove (\ref{4.4}), it remains to
show that the oscillatory integrals
in (\ref{4.6}) converge to zero.
Let us first analyze
the entries of the matrix $q^{ij}_0$, $i,j=1,...,8$.

\begin{lemma} \la{l4.1}
Let the assumptions of  Theorem~A hold.
Then $\hat q_0^{ij}\in L^1(\R^3)$ for any $i,j$.
\end{lemma}
{\bf Proof.}
Let us first  prove that
\beqn
\la{4.7}
q^{ij}_0(z)&\in& L^p(\R^3),\,\,\,\,p\ge 1,
\quad i,j=1,...,8.
\eeqn
Conditions
{\bf S0}, {\bf S2} and {\bf S3} imply
by (cf. \ci[Lemma 17.2.3]{IL}) that
\be\la{4.9}
|q^{ij}_0(z)|\le
Ce_0\varphi^{1/2}(|z|),~~~
z\in\R^3,~~~i,j=1,\dots,8.
\ee
The mixing coefficient $\varphi$ is bounded, and hence
relations  (\ref{4.9}) and (\ref{1.12}) imply
(\ref{4.7}),
$$
\int\limits_{\R^3} |q^{ij}_0(z)|^p\,dz\le
Ce_0^p\int\limits_{\R^3}
\varphi^{p/2}(|z|)\,dz\le  C_1
 \int\limits_0^\infty r^2 \varphi^{1/2}(r)\,dr <\infty.
$$
By Bohner's theorem, $\hat q^{ij}_0$
is a nonnegative matrix-valued measure
on $\R^3$, and condition
{\bf S2} implies that the total measure
 $\hat q_0(\R^3)$ is finite.
On the other hand, relation (\ref{4.7}) for $p=2$
gives $\hat q^{ij}_0\in  L^2(\R^3)$.
Hence, $\hat q^{ij}_0\in  L^1(\R^3)$.
\hfill$\Box$\\

Let us apply this lemma to the oscillatory integrals 
entering  (\ref{4.6}).
The convergence (\ref{4.4}) follows from (\ref{4.6})
by the Lebesgue-Riemann theorem.
 This completes the proof of  Proposition \ref{p4.1}.
\hfill$\Box$\\

Relation (\ref{4.7}) for $p=1$
implies now that $\hat q_0(k)$
 is bounded. Hence, the explicit formula
 (\ref{1.13}) implies the following assertion.
\begin{cor}\la{coro}
All matrix elements $\hat q^{ij}_\infty(k)$,
 $i,j=1,\dots,8$, are bounded. 
\end{cor}

\setcounter{equation}{0}
\section{Convergence of characteristic functionals}

To prove  Proposition \ref{l2.2}
for the general case of a non-Gaussian measure $\mu_0$,
 we  develop a version of Bernstein's ``room - corridor''
method of  \ci{DKKS}-\ci{DKS1}:
(i) we use an integral
representation  for the solutions of (\ref{21'}),
 (ii) divide the region of the integration
into "rooms" and "corridors" and
(iii) evaluate their contribution.
As the result, the value $\langle U(t)\psi_0,\phi\rangle$
for $\phi\in{\cal D}$ is represented
as the sum of weakly dependent random variables.
Then we  apply Bernstein's ``rooms-corridor'' method and
 the Lindeberg central limit theorem.

(i)  We first evaluate the inner product
$\langle U(t)\psi_0,\phi\rangle$ in (\ref{2.6})
by using  duality arguments.
For  $t\in\R$, introduce a ``formal adjoint'' operators
 $U'(t)$
from the space ${\cal D}$ to a suitable space of distributions.
For example,
\be\la{def0}
\langle \psi,U'(t)\phi\rangle =
\langle U(t)\psi,\phi\rangle ,\,\,\,
\phi\in {\cal D},
\,\,\, \psi\in {\cal H}.
\ee
 Write $\phi(\cdot,t)=U'(t)\phi$. Then
(\ref{def0}) can be represented as
\be\la{defY}
\langle \psi(t),\phi\rangle =
\langle \psi_0,\phi(\cdot,t)\rangle,
\,\,\,\,t\in\R.
\ee
The adjoint groups
admit a convenient description
(see Lemma \re{ldu} for the group $U'(t)$).
\begin{lemma}\la{ldu}
For $\phi\in {\cal D}$, the function
$U'(t)\phi=\phi(x,t)$  is the solution of
\be\la{UP0}
\dot\phi(x,t)=(\alpha\cdot\nabla+i\beta m)\phi(x,t),~~~\phi(x,0)=\phi(x).
\ee
\end{lemma}
{\bf Proof.}
Differentiating (\ref{def0}) with respect to $t$ for
$\psi,\phi\in {\cal D}$, we obtain
\be\la{UY0}
\langle\psi,\dot U'(t)\phi\rangle=
\langle\dot U(t)\psi,\phi\rangle.
\ee
The group $U(t)$ has the generator
$
{\cal A}= -\alpha\cdot\nabla-i\beta m.
$
Therefore,
 the generator of $U'(t)$   is the conjugate operator
\be\la{A0'}
{\cal A}'=\alpha\cdot\nabla+i\beta m.
\ee
Hence, relation (\ref{UP0}) holds indeed with
 $\dot \phi={\cal A}'\phi$.
 \hfill$\Box$
\begin{remark}\la{convU0}
Comparing (\ref{UP0}) and (\ref{21'}),
we see that $\phi(x,t)=U'(t)\phi$ can be represented
as a convolution (cf (\ref{25})), namely,
\be\la{conr}
\phi(\cdot,t)={\cal R}_t*\phi,\quad
{\cal R}_t:
=(\partial_t +\alpha\cdot\nabla+im\beta) E_t.
\ee
\end{remark}

(ii) Introduce a ``room-corridor'' partition of
$\R^3$. For a given $t>0$, choose $d_t\ge 1$ and
$\rho_t>0$ such that
$\rho_t\sim t^{1-\delta}$ with some $\de\in (0,1)$
and $d_t\sim t/{\ln t}$, as  $t\to\infty.$
Set $h_t=d_t+\rho_t$ and
\be\la{rom}
a^j=jh_t,\,\,\,b^j=a^j+d_t,
\,\,\,j\in\Z.
\ee
We refer to the slabs
$R_t^j=\{x\in\R^3:~a^j\le x^3\le b^j\}$ as ``{\it rooms}''
and to $C_t^j=\{x\in\R^3:~b^j\le x^3\le  a_{j+1}\}$ as
 ``{\it corridors}''.
Here  $x=(x^1,x^2,x^3)$, the symbol $d_t$ stands for
 the {\it width} of a room, and
$\rho_t$  for that of a corridor.

Denote  by
 $\chi_r$ the indicator of the interval $[0, d_t]$ and
by $\chi_c$ that of $[d_t, h_t]$,
which means that
$$
\sum_{j\in \Z}(\chi_r(s-jh)+\chi_c(s-jh))=1
\quad \mbox{for (almost all) }\,\, s\in\R.
$$
The  following  decomposition holds:
\be\la{res}
\langle \psi_0,\phi(\cdot,t)\rangle =
 \sum_{j\in \Z}
(\langle \psi_0,\chi_r^j\phi(\cdot,t)\rangle +
\langle
\psi_0,\chi_c^j\phi(\cdot,t)\rangle ),
\ee
where $\chi_r^j:=\chi_r(x^3-jh)$ and
$\chi_c^j:=\chi_c(x^3-jh)$.
Consider the random variables
 $ r_t^j$ and $ c_t^j$ given by
\be\la{100}
r_t^j= \langle \psi_0,\chi_r^j\phi(\cdot,t)\rangle  ,~~
c_t^j= \langle \psi_0,\chi_c^j\phi(\cdot,t)\rangle ,~~~~~~j\in\Z.
\ee
Then  (\ref{res}) and (\ref{defY}) imply
\be\la{razli}
\langle U(t)\psi_0,\phi\rangle =\sum\limits_{j\in\Z}
(r_t^j+c_t^j).
\ee
The series  in (\ref{razli})
is in fact a finite sum.
 Indeed, for the support  of
$\phi$ we have 
$$
{\rm supp}\,\phi\subset B_{\ov r}\quad
\mbox{ for some }\,\,{\ov r}>0.
$$
Then, by  the  convolution representation (\ref{conr}),
the support of the function $\phi(\cdot,t)$ at $t> 0$
is a subset of an  ``inflated future cone''
\be\la{conp}
{\rm supp}\hspace{0.5mm}
\phi\subset\{(x,t)\in\R^3\times{\R_+}:
~ |x|\le t+{\ov r}\},
\ee
whereas ${\cal R}_t(x)$ is supported by the `future cone' $|x|\le t$.
The latter fact follows from (\ref{conr})
and  from Remark \ref{r2.1}.
Finally, it follows from (\ref{100})   that
\be\la{1000}
r_t^j= c_t^j= 0
\,\,\,\quad{\rm  for }\quad\quad \,jh_t+t< -{\ov r}\,\,\,
\quad\mbox{and for }\quad\,\,jh_t-t>{\ov r}.
\ee
Therefore, the series (\re{razli}) becomes
a  sum,
\be\la{razl}
\langle U(t)\psi_0,\phi\rangle =
\sum\limits_{-N_t}^{N_t}
(r_t^j+c_t^j),\,\,\,\,\ds N_t\sim \fr th_t.
\ee
\begin{lemma}  \la{l5.1}
    Let  Conditions {\bf S0--S3} hold.
Then the following bounds hold for $t>1$:
\be\la{106}
E|r^j_t|^2\le  C(\phi)~d_t/t,\,\,\,\,
E|c^j_t|^2\le C(\phi)~\rho_t/t,\,\,\,\,j\in\Z.
\ee
\end{lemma}
{\bf Proof.}
We discuss the first bound in (\ref{106}) only, 
because the other can be proved in a similar way.
Rewrite the LHS of
(\ref{106}) as the integral of correlation functions.
We  obtain
 \be\la{100rq}
E|r_t^j|^2= \langle\chi_r^j(x_3)\chi_r^j(y_3)q_0(x-y),
\phi(x,t)\otimes\phi(y,t)\rangle.
\ee
The following uniform bound holds
(cf. \ci[Thm. XI.17 (b)]{RS3}):
\be\la{bphi}
\sup_{x\in\R^3}|\phi(x,t)| ={\cal O}(t^{-3/2}),
\,\,\,\,t\to\infty.
\ee
In fact, (\ref{conr}) and  (\ref{fs}) imply that
the function $\phi(x,t)$ can be represented written 
as the sum
\be\la{frepe}
\phi(x,t)=\sum\limits_{\pm}
\int\limits_{\R^3} e^{-i(kx\pm\om t)}
a^\pm(\om)
\hat\phi( k)~d k,
\ee
where $a^\pm(\om)$ is a matrix whose entries
are linear functions of $\om$ or $1/\om$.
Let us prove the asymptotics (\re{bphi}) along each ray
$x=vt+x_0$ with   $|v|\le 1$.
The asymptotic relation thus obtained must hold
 uniformly in
$x\in\R^3$ by (\ref{conp}). By (\ref{frepe}) we have
\be\la{freper}
\phi(vt+x_0,t)=\sum\limits_{\pm}
\int\limits_{\R^3} e^{-i(kv\pm\om)t-ikx_0}
a^\pm(\om) \hat\phi( k)~d k.
\ee
This is a sum of oscillatory integrals with the phase
functions $\phi_\pm(k)=kv\pm\om(k)$. Each function
 has two stationary points which are
solutions of the equation $v=\mp\nabla\om(k)$
if $|v|<1$, and has none if $|v|\ge 1$.
The phase  functions are nondegenerate, i.e.,
\be\la{Hess}
{\rm det}\left(\ds\frac{\pa^2\phi_\pm(k)}{\pa k_i\pa k_j}\right)_{i,j=1}^3
\ne 0, \quad k\in\R^3.
\ee
Finally, $\hat\phi(k)$ is smooth and  rapidly decays
at infinity.
Therefore, $\phi(vt+x_0,t)={\cal O}(t^{-3/2})$
according  to the standard method of
stationary phase, see \ci{F}.

 According to
(\ref{conp})  and (\ref{bphi}),
 it follows from  (\ref{100rq}) that
\be\la{er}
E|r_t^j|^2\le Ct^{-3}\int\limits_{|x|\le t+{\ov r}}
\chi_r^j(x^3)\Vert q_0(x-y)\Vert ~dx dy
= Ct^{-3}\int\limits_{|x|\le t+{\ov r}}\!\!
\chi_r^j(x^3)dx~
\int\limits_{\R^3}\!\!
\Vert q_0(z)\Vert dz,
\ee
where $\Vert q_0(z)\Vert $ stands for  the norm of the
 matrix $\left(q_0^{ij}(z)\right)$.
Therefore, relation  (\ref{106})  follows for
$\Vert q_0(\cdot)\Vert \in L^1(\R^3)$ by
(\ref{4.7}).
\hfill$\Box$
\medskip

Hence, the rest of the
proof of Proposition \ref{l2.2}
is the same as that in the case
 of the Klein-Gordon equation,  \cite[p.20-25]{DKKS}.
The proof of Theorem~A  is complete.


\end{document}